\documentclass[11pt,letterpaper]{article}
\usepackage[T1]{fontenc}
\usepackage[utf8]{inputenc}
\usepackage[margin=1in]{geometry}
\usepackage{mathptmx}
\usepackage{graphicx}
\usepackage{booktabs}
\usepackage{longtable}
\usepackage{array}
\usepackage{tabularx}
\usepackage{float}
\usepackage{caption}
\usepackage{amsmath,amssymb}
\usepackage{hyperref}
\usepackage{setspace}
\usepackage{titlesec}
\usepackage{enumitem}
\titleformat{\section}{\normalfont\Large\bfseries}{\thesection.}{0.5em}{}
\titleformat{name=\section,numberless}{\normalfont\Large\bfseries}{}{0em}{}
\hypersetup{hidelinks}
\begin{document}

\begin{center}
{\LARGE \textbf{The HyperFrog Cryptosystem: Connected Graph Cycle Rank \\ as a Structured Secret Distribution for an Experimental Post \\ Quantum KEM}}\\[1.2em]
Victor Duarte Melo\\
Independent Researcher\\
Erdos Number 4\\
\texttt{victormeloasm@gmail.com}\\[0.8em]
Revised manuscript for IACR style circulation\\
March 2026, implementation aligned with HyperFrog v36.0 and benchmark schema v3
\end{center}

\begin{abstract}
HyperFrog is an experimental post quantum key encapsulation mechanism whose public key layer follows an unstructured Learning With Errors design while the secret vector is sampled from a constrained family of voxel occupancy fields. Earlier drafts mixed three different ideas too loosely, namely an informal topological narrative, an engineering miner kept for convenience in code, and a formal miner intended to define the claimed secret law. The present revision separates these pieces explicitly.

The formal miner is now stated as a connected frontier growth process on the $16 \times 16 \times 16$ voxel grid. Starting from a uniformly chosen seed voxel, the miner repeatedly adds a uniformly chosen frontier voxel until the occupied graph has exact weight 2048. A user supplied graph cycle rank threshold $\tau$ may then be applied to the resulting connected exact weight shape. The practical miner remains in code for experimentation, but it is excluded from formal claims. This makes the secret law explicit and keeps the paper honest about which implementation mode is being analyzed.

The algebraic core remains intentionally conservative. The public key is an LWE instance over $q = 2^{16}$ with $N = 4096$, $M = 2048$, and $k = 256$ encapsulated bits. Matrix rows are expanded from a short seed by a stream cipher, secrets are binary, noise is centered binomial, and chosen ciphertext security is pursued through a Fujisaki--Okamoto style transform in the random oracle model. The novelty lies in secret distribution engineering rather than in a new hardness assumption. No reduction is claimed from the connected graph exact weight secret law to standard binary secret LWE.

This revision also aligns the manuscript with the current implementation and benchmark schema. The code now exposes formal and practical miner modes with sharply separated semantics, records per sample mining diagnostics, records decapsulation correctness diagnostics, and splits file decryption into secret key unlock time, cryptographic decryption time, and total end to end time. Across the uploaded formal benchmark corpus used in this manuscript, the implementation produced 1670 formal key generation samples with 1670 successes, zero timeouts, exact occupied weight 2048 in every formal sample, one connected component in every formal sample, and 1670 out of 1670 valid decapsulation diagnostics. The observed accepted cycle ranks in those formal samples ranged from 2315 to 2856 with mean 2687.3.

The same corpus also clarifies what can and cannot yet be claimed from the benchmarks. In the current artifact suite, most throughput runs use permissive target cycle rank thresholds such as 8, while the observed accepted cycle ranks are already around $2.6 \times 10^3$ to $2.8 \times 10^3$. Consequently, the present threshold and budget sweeps characterize implementation stability more than they characterize a hard acceptance regime. Likewise, the current JSON field named \texttt{acceptance\_probability} is an instrumentation ratio inside the parallel search loop and should not yet be interpreted as a calibrated acceptance probability of the formal law itself.

The practical conclusion is therefore bounded and empirical. HyperFrog is a transparent research system whose formal secret law is now stated much more clearly, whose benchmark semantics are much less ambiguous than in earlier drafts, and whose present artifact suite supports stable implementation claims. It does not yet provide a reduction for the structured secret family, a calibrated acceptance study for difficult thresholds, or a deployment level security argument. The purpose of this manuscript is to state the construction precisely, to separate formal and engineering claims cleanly, and to document the current empirical behavior of the implementation in a form suitable for external review.

\end{abstract}

\begin{center}
HyperFrog revision aligned with v36.0 and benchmark schema v3
\end{center}

\section{Introduction}
The migration toward post quantum public key cryptography has produced a mature family of KEMs built from lattices, code based systems, and isogenies, together with a large ecosystem of proofs, implementations, and attacks. Within that landscape it is still scientifically useful to study experimental constructions that change only one part of the design at a time. HyperFrog belongs to that exploratory line of work. Its algebraic core is intentionally familiar, but its secret distribution is not. Instead of drawing an unconstrained binary secret, the construction mines a voxel occupancy field whose occupied six neighbor graph has a large cycle rank. This voxel based viewpoint is also compatible with the scale normalized geometric framework for measurable shapes developed in~\cite{melo2025vffnd}.

The original manuscript already made one conservative point correctly. The security target of HyperFrog is not a stand alone topological problem. The intended hardness claim is still an LWE style claim. The role of topology is to define a structured secret distribution and a distinctive interpretation of the secret key. That separation remains the right way to think about the scheme and this revision keeps it explicit throughout.

Two issues nevertheless required a substantial rewrite. The first issue is terminological. Earlier text used the word genus as a mnemonic for many loops. In the actual code and proofs, however, the mined quantity is $\beta_1$ of the occupied adjacency graph, namely the graph cycle rank $|E| - |V| + C$. That quantity is mathematically precise and useful, but it is not the same thing as the genus of a three dimensional body. The revised paper therefore uses the formal language of graph cycle rank and graph Betti number everywhere, while acknowledging that the old mnemonic was informal.

The second issue is more important. Earlier drafts described a formal rejection sampler that accepted a uniformly sampled occupancy field if it passed a wide popcount window and a very small cycle rank threshold. At the published parameters that predicate was too weak. In practical terms it left the secret law extremely close to an unconstrained binary source. The present revision does not defend that rule. Instead it defines a new formal miner whose law is explicit, connected by construction, exact in its occupied weight, and optionally conditioned by a tunable cycle rank threshold.

This shift changes the role of the mining section. It is no longer a loose intuition pump about topological rejection sampling. It becomes a formal description of the secret law actually used in the revised implementation. At the same time the engineering miner remains available in code, but only as an implementation mode for experimentation, not as the subject of formal claims. This separation is central to the revision and is one of the main reasons for reissuing the manuscript in expanded form.

The paper is organized as follows. Section 2 summarizes the revision at a high level. Section 3 states the overall design and parameters. Sections 4 and 5 formalize the graph model and the revised connected graph miner. Sections 6 and 7 specify the KEM and its correctness properties. Section 8 discusses security assumptions and limitations. Sections 9 through 12 document the implementation and benchmarks, including the new split between secret key unlock cost and cryptographic file processing cost. The final sections discuss limitations, open problems, and future work.

\section{Revision summary and statement of changes}
Because this document is intended as a revision rather than a completely unrelated replacement, it is useful to state the changes in one place before the technical sections begin. The table below compares the earlier manuscript with the revised one and identifies the points that were rewritten in order to align the text with the current implementation and benchmark corpus.

\begin{table}[H]
\centering
\caption{Summary of the main manuscript level corrections introduced in the v36 aligned revision.}
\renewcommand{\arraystretch}{1.25}
\begin{tabularx}{\textwidth}{>{\raggedright\arraybackslash}p{0.20\textwidth} >{\raggedright\arraybackslash}p{0.37\textwidth} >{\raggedright\arraybackslash}X}
\toprule
\textbf{Topic} & \textbf{Earlier manuscript} & \textbf{Revised manuscript} \\
\midrule
Topological invariant & Informal genus language was used broadly. & The formal invariant is $\beta_1$ of the occupied six neighbor graph, also called graph cycle rank. The text now uses this terminology consistently. \\
Formal miner & Uniform occupancy sampling with a permissive popcount window and a weak rejection rule was described as the formal path. & The formal miner is now a connected frontier growth process with exact weight 2048 and a tunable cycle rank threshold $\tau$. The benchmark corpus documents the specific $\tau$ values actually used. \\
Connectivity & Connectivity appeared as optional or implementation dependent. & Connectivity is mandatory in the formal miner because the growth process builds one connected component by construction. \\
Weight control & The old rule used a wide window between 1200 and 2600. & The revised formal miner fixes the occupied weight exactly at 2048, which makes both the distribution and the noise analysis cleaner. \\
Practical miner & The engineering path and the formal path were not sharply separated. & The practical miner is explicitly classified as an engineering mode for experimentation only. It is excluded from formal statements. \\
Benchmark semantics & File decryption time mixed password unlock with cryptographic processing and key generation statistics were too coarse. & The benchmark now reports unlock time, cryptographic decryption time, total time, per sample key generation diagnostics, and decapsulation correctness diagnostics. \\
Scope of benchmark claims & Single run timing numbers were easy to overread as stronger evidence than they were. & The present manuscript treats the uploaded artifacts as an implementation study. It states explicitly which sweeps are informative and which are still non binding. \\
\bottomrule
\end{tabularx}
\end{table}

These changes are not cosmetic. They alter both the meaning of the mining section and the interpretation of the experimental results. In the earlier text, the formal and practical notions of mining were too close together. In the revised manuscript the formal rule is explicit, narrow, and reproducible. At the same time the benchmark corpus makes clear that the currently uploaded throughput runs mostly use permissive thresholds. The paper therefore no longer treats threshold filtering as empirically calibrated just because the code supports it.

The practical miner remains available in code because a prototype often needs fast smoke test modes and timeout tolerant engineering paths. The revised paper documents that path honestly, labels it as practical rather than formal, and uses it only in comparison experiments.

The benchmark rewrite is equally important. The previous end to end file decryption number folded together two conceptually different costs, namely password based secret key unlock and the cryptographic file path driven by the KEM shared secret. The current schema separates those measurements and also records mining statistics and decapsulation diagnostics. This produces a much cleaner artifact, but it also makes one limitation more visible: the current JSON field named \texttt{acceptance\_probability} is an implementation ratio inside the parallel search loop, not yet a calibrated scientific acceptance probability for the formal law itself.

This revised manuscript therefore has a narrower but cleaner claim. HyperFrog is an experimental unstructured LWE KEM with a connected growth exact weight secret generator, an optional cycle rank threshold, and an implementation study whose artifact semantics are stated explicitly.

\section{Design overview and parameter set}
HyperFrog has three conceptual layers. The first layer is a secret generator on the $16 \times 16 \times 16$ voxel grid. The second layer is an unstructured LWE public key built from that secret. The third layer is an FO style transform that turns the CPA core into a random oracle model KEM. The key point of the design is that only the secret law is unusual. The matrix expansion, public key equation, and encapsulation pattern all remain within the broad family of lattice based KEM ideas.

The implementation constants in \texttt{hyperfrog36.cpp} are $N = 4096$, $M = 2048$, $q = 65536$, and \texttt{KBITS} $= 256$. The voxel secret therefore occupies the full 4096 bit occupancy field of a $16 \times 16 \times 16$ lattice. The public key is serialized as a 32 byte seed for matrix expansion together with $M$ sixteen bit coefficients of $b$. This yields $32 + 2M = 4132$ bytes before small file level headers. The current ciphertext remains intentionally large because the scheme stores one $u$ vector per encapsulated bit. With $k = 256$ and $N = 4096$, the dominant term is $2kN$ bytes, so the ciphertext is approximately 2.10 MB.

The formal miner uses exact occupied weight 2048 together with a user supplied cycle rank threshold $\tau$. No single threshold is treated as canonical in this manuscript. The uploaded benchmark corpus uses $\tau = 8$ in most throughput runs, $\tau \in \{6,8,10,12\}$ in a small threshold sweep, and mining budgets between 1000 ms and 30000 ms in a budget sweep. Because the observed accepted cycle ranks are already much larger than those threshold values, these runs should be read as implementation benchmarks and not as a calibrated acceptance study.

The encapsulation layer uses a row subset mechanism of weight 64 per message bit. For each of the 256 message bits, the code derives a deterministic seed, uses it to sample 64 matrix rows, sums those rows to obtain one $u$ vector, and adds a small centered binomial perturbation. The corresponding $v$ coordinate is computed by summing the matching public key entries, adding a small scalar noise term, and then adding $q/2$ when the message bit is one.

The file encryption utility is built on top of the KEM shared secret. A symmetric key is derived by SHA3 based domain separated hashing, and the implementation uses AES 256 GCM for file encryption and decryption. Secret key files can be left unprotected or wrapped under a password. When password protection is enabled, the code derives a wrapping key by Argon2id and authenticates the stored blob with AEAD. This detail matters for benchmark interpretation because password based secret key unlock can dominate end to end decryption latency even when the cryptographic file path itself is fast.

\begin{table}[H]
\centering
\caption{Main reference parameters of the v36 aligned implementation and benchmark corpus.}
\renewcommand{\arraystretch}{1.2}
\begin{tabularx}{\textwidth}{>{\raggedright\arraybackslash}p{0.30\textwidth} >{\centering\arraybackslash}p{0.17\textwidth} >{\raggedright\arraybackslash}X}
\toprule
\textbf{Parameter} & \textbf{Value} & \textbf{Comment} \\
\midrule
$N$ & 4096 & Voxel occupancy vector length, corresponding to a $16 \times 16 \times 16$ lattice. \\
$M$ & 2048 & Number of public key rows and $b$ coefficients. \\
$q$ & 65536 & Modulus for all LWE arithmetic. \\
\texttt{KBITS} & 256 & Number of encapsulated bits and shared key derivation width. \\
Formal occupied weight & 2048 & Exact occupied voxel count in the formal miner. \\
Cycle rank threshold $\tau$ & User supplied & The benchmark corpus uses 6, 8, 10, and 12. No single value is treated as canonical in this manuscript. \\
Row subset weight & 64 & Number of selected rows used for each encapsulated bit. \\
Public key size & 4132 bytes & 32 byte seed plus 2048 sixteen bit coefficients. \\
Ciphertext size & About 2.10 MB & Dominated by 256 vectors of length 4096 with sixteen bit coefficients. \\
Benchmark schema & v3 & Includes per sample miner diagnostics and decapsulation correctness diagnostics. \\
\bottomrule
\end{tabularx}
\end{table}

\section{Graph model and topological statistic}
Let $G$ be the 16 by 16 by 16 voxel grid. A candidate secret shape $S$ is a subset of $G$ and is identified with its binary occupancy vector $s \in \{0,1\}^{4096}$ under a fixed flattening order. The revised manuscript formalizes the topological statistic entirely in graph theoretic terms. This avoids the ambiguity that comes from speaking about genus for a three dimensional voxel body without specifying a boundary extraction method and a homology convention.

Given $S$, define $\Gamma(S)$ to be the undirected six neighbor graph whose vertex set is the occupied voxels and whose edges connect face adjacent occupied voxels. If $V$ denotes the number of occupied voxels, $E$ denotes the number of occupied face adjacencies, and $C$ denotes the number of connected components, then the graph cycle rank is defined by
\[
\beta_1(\Gamma(S)) = E - V + C.
\]
This is the standard first Betti number of a finite graph, or equivalently the dimension of its cycle space over $\mathbb{F}_2$.

This invariant is mathematically precise, computationally inexpensive, and directly aligned with the implementation. The topology routine in \texttt{hyperfrog36.cpp} computes $V$ by population count over the bitset, computes $E$ by scanning positive $x$, $y$, and $z$ directions only, and computes $C$ by breadth first search over occupied voxels. The reported quantity named \texttt{cycle\_rank} in the code is exactly $E - V + C$.

The revised paper therefore avoids two overstatements. First, it does not identify graph cycle rank with the genus of the underlying voxel body. Second, it does not imply that a large graph cycle rank necessarily yields a topologically complicated boundary surface in any specific geometric sense. What it does guarantee is a large number of independent cycles in the occupied adjacency graph.

This distinction matters because the miner conditions on the graph invariant, not on a continuous surface invariant. The graph viewpoint is nevertheless rich enough to give the scheme a structured secret law and to support exact implementation. In an experimental cryptosystem that already relies on a large binary occupancy field, graph cycle rank is a reasonable and honest topological statistic to mine on.

For a connected shape, $C = 1$, so the formula simplifies to $\beta_1(\Gamma(S)) = E - V + 1$. Because the formal miner enforces connectedness by construction, this is the effective form used by the acceptance rule.

\section{Formal connected graph miner}
The central technical correction in the v36 aligned branch is the replacement of the earlier weak rejection sampler by a connected frontier growth process. The revised miner does not begin by filling the full lattice with independent Bernoulli bits and then asking whether the result accidentally has the desired properties. Instead it constructs a connected occupied graph one voxel at a time.

The process begins by selecting one start voxel uniformly from the 4096 grid positions. That voxel is marked occupied, and its currently unoccupied six neighbors are inserted into a frontier set. At each subsequent step the miner chooses one frontier voxel uniformly at random, adds it to the occupied set, and then updates the frontier by inserting any newly exposed unoccupied neighbors. The process stops when the occupied set has exact size 2048.

This construction has three useful properties. First, connectivity is guaranteed from the start, because every newly added voxel touches the existing occupied component. Second, the occupied weight is exact rather than approximate. Third, the random growth law remains explicit. It is not the conditional law of an ambient product measure. It is a specific sequential distribution over connected exact weight voxel graphs generated by random frontier expansion.

Once a candidate shape of weight 2048 has been built, the implementation computes its graph cycle rank and accepts only if $\beta_1$ is at least a user supplied threshold $\tau$. If the time budget expires before a successful shape is found, the formal miner fails. There is no deterministic fallback in this path, because a fallback would define a different output law and would undermine the claim that the formal miner samples from the stated distribution.

The code contains a separate practical miner for engineering use. That path begins with a random bitset, keeps the largest connected component, and can fall back to a deterministic connected shape on timeout. Those behaviors are useful for implementation testing, but they do not belong in the formal specification. The present paper therefore names that path explicitly as the engineering miner and excludes it from formal claims.

The induced secret law of the formal miner can be written as follows. Let $G_{\mathrm{conn},2048}$ be the set of connected voxel graphs on the $16 \times 16 \times 16$ lattice with exactly 2048 occupied vertices. Let $P_{\mathrm{grow}}$ be the distribution induced by the frontier growth process on this set. For a chosen threshold $\tau$, the formal secret distribution is the conditional law
\[
P_{\mathrm{formal},\tau} = P_{\mathrm{grow}} \mid \beta_1 \ge \tau.
\]
This is a structured and non uniform binary secret law by design. The paper does not attempt to rewrite it as a near uniform source, because that would obscure rather than clarify the actual construction.

One practical consequence is that entropy arguments must be handled carefully. In the earlier rejection picture, one could at least discuss acceptance probability relative to a uniform ambient source. In the connected growth picture the base law is already non uniform. The relevant question is no longer closeness to the full cube $\{0,1\}^{4096}$. It is whether the resulting structured binary secret family remains a plausible small secret family for unstructured LWE, and whether any of its graph induced correlations lead to attacks. That remains open.

The benchmark corpus supplied with this revision should be read with one extra caveat. The throughput suites use permissive thresholds, mostly $\tau = 8$, while the observed accepted cycle ranks already cluster around $2.6 \times 10^3$ to $2.8 \times 10^3$. This means that the present artifact suite primarily characterizes the connected growth exact weight law and the implementation around it. It does not yet isolate a difficult acceptance regime in which threshold filtering is the dominant event.

\begin{quote}\small
\textbf{Algorithm box 1.} Pseudocode for the formal miner used by the revised specification.

\medskip
\textbf{Algorithm 1. Formal connected graph miner}

\textit{Input:} target weight $W = 2048$, threshold $\tau$, time budget $T$
\begin{enumerate}[leftmargin=2em,label=\arabic*.]
\item Choose one start voxel uniformly from the 4096 grid points.
\item Mark it occupied and initialize the frontier with its unoccupied six neighbors.
\item While the occupied set has fewer than $W$ voxels:
  \begin{enumerate}[leftmargin=2em,label={}]
  \item choose one frontier voxel uniformly at random,
  \item add it to the occupied set,
  \item update the frontier with any newly exposed unoccupied neighbors.
  \end{enumerate}
\item Compute $\beta_1(\Gamma(S)) = E - V + 1$.
\item Accept if $\beta_1(\Gamma(S)) \ge \tau$. Otherwise restart until time budget $T$ expires.
\item On timeout, return failure. No fallback is used in the formal path.
\end{enumerate}
\end{quote}

The miner is still computationally straightforward. It uses only random frontier selection, bitset updates, a breadth first search based connected component count, and edge counting. This simplicity is useful for a research prototype because it makes the secret law explicit without introducing difficult geometric preprocessing.

\begin{figure}[H]
\centering
\includegraphics[width=0.78\textwidth]{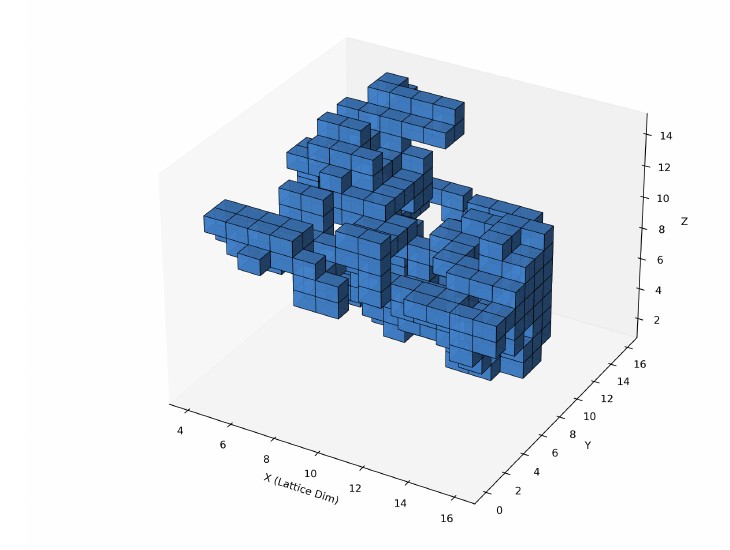}
\caption{Example voxel rendering used as a qualitative illustration of the kind of connected occupancy field produced by the formal connected growth miner. The formal invariant mined by the implementation is graph cycle rank $\beta_1$ of the occupied six neighbor graph.}
\end{figure}

\section{Formal and engineering miner separation}
The presence of two miner modes in the code deserves its own section because it is a common source of confusion in prototype cryptography. The formal miner defines the construction claimed by the paper. The engineering miner is a convenience mode for development, smoke testing, and practical experimentation. Conflating the two would produce a mismatch between the described secret law and the implemented output distribution.

The engineering miner in \texttt{hyperfrog36.cpp} begins from a random bitset, retains only the largest connected component, and searches for a high cycle rank object under a time budget. On timeout it may return a deterministic connected fallback shape. This is acceptable for testing file wrappers, parser logic, and end to end encryption loops, but it is not an admissible formal secret generator for the purposes of this manuscript.

The revised text therefore enforces a strict rule for interpretation. Every theorem, definition, and benchmark claim tied to the formal construction refers to the connected frontier growth miner and its exact weight high threshold acceptance rule. Every statement about the practical miner is labeled as engineering only. This is especially important in the benchmark section, because a practical mode can make key generation appear more stable at the cost of changing the secret distribution fundamentally.

This separation also improves future research. Because the engineering mode remains in the code, one can use it as a laboratory for heuristic secret generation strategies, such as aggressive normalization, richer geometric filters, or different fallback logic. None of those experiments need to contaminate the formal statement of the current scheme.

\section{HyperFrog KEM construction}
The revised implementation keeps the LWE style core deliberately close to the earlier design. A 32 byte seed expands the public matrix row by row through ChaCha20 in IETF mode. For each row index $i$, the nonce is derived from $i$ and the stream output is parsed into $N$ sixteen bit words. This yields an implicit matrix $A \in \mathbb{Z}_q^{M \times N}$ without storing the full dense matrix.

Key generation first runs the formal miner to obtain a binary secret $s \in \{0,1\}^{4096}$. It then samples noise $e \in \mathbb{Z}^M$ from a centered binomial distribution with parameter 4 and computes $b = As + e \bmod q$. The public key is $(\mathrm{seed}_A, b)$, while the secret key stores the shape together with the public key so that FO style reencryption can be performed during decapsulation.

Encapsulation begins by sampling a 256 bit message string $\mu$. The code computes a public key digest and derives a deterministic row subset seed from $\mu$ and that digest. For each bit position $j$, the implementation samples 64 row indices, expands the corresponding rows, sums them to form one $u^{(j)}$ vector, adds centered binomial noise to that vector, and computes the associated scalar coordinate $v_j$ from the matching $b$ entries and one additional scalar noise term. The bit $\mu_j$ is embedded by adding $q/2$ when $\mu_j = 1$.

The ciphertext therefore contains 256 vectors of length 4096, stored in transposed form for efficient decapsulation, together with 256 scalar coordinates and a transcript binding tag. The current scheme is intentionally not size optimized. The purpose of the present implementation is to study the interaction between a structured binary secret and an unstructured LWE style KEM, not to compete with compact standardized designs on bandwidth.

Decapsulation recomputes each message bit by forming $t_j = v_j - \langle u^{(j)}, s \rangle \bmod q$ and applying a branchless $q/4$ threshold test that decides whether $t_j$ is closer to $q/2$ than to zero. The candidate $\mu'$ is then used to rederive the deterministic row subset seed, reconstruct the expected ciphertext, recompute the tag, and finally choose between a real and fake shared key in constant time. This is a standard FO style pattern adapted to the present sparse row subset setting.

For file encryption the KEM shared secret is hashed into an AES 256 GCM key. The benchmarked utility uses streaming file processing and reports file path costs independently from KEM costs. Because the secret key file can itself be password protected, decryption may involve a large Argon2id based unlock phase before the actual file decryption begins. The benchmark methodology now reports that step separately.

\begin{align*}
b &= As + e \bmod q, \\
u^{(j)} &= \sum_{i \in I_j} A_i + e_1^{(j)} \bmod q, \qquad
v_j = \sum_{i \in I_j} b_i + e_2^{(j)} + \left\lfloor \frac{q}{2} \right\rfloor \mu_j \bmod q, \\
t_j &= v_j - \langle u^{(j)}, s \rangle \bmod q.
\end{align*}

\section{Correctness analysis}
The correctness argument follows the same general structure as in the earlier manuscript, but the revised miner makes one part cleaner because the occupied weight is fixed exactly at 2048. Let $\eta_j$ denote the total noise term appearing after cancellation of the noiseless inner products in bit position $j$. The decoder succeeds whenever $|\eta_j|$ is strictly below $q/4$, because the centered representative of $t_j$ is then much closer to zero when $\mu_j = 0$ and much closer to $q/2$ when $\mu_j = 1$.

Writing the full relation gives $t_j = \lfloor q/2 \rfloor \mu_j + \eta_j \bmod q$, where $\eta_j$ contains three contributions: the sum of the 64 public key noise terms $e_i$ chosen by the row subset $I_j$, one scalar noise term $e_2^{(j)}$, and the negative inner product of the vector noise $e_1^{(j)}$ with the binary secret $s$. The code uses a centered binomial source with variance 2 for each scalar sample.

Because the row subset weight is 64, the variance of the summed public key noise contribution is $64 \times 2 = 128$. The scalar term contributes variance 2. The inner product term contains exactly $\mathrm{wt}(s)$ centered binomial samples because the secret is binary, and in the revised formal miner $\mathrm{wt}(s) = 2048$ exactly. Therefore its variance is $2048 \times 2 = 4096$. Summing the independent contributions yields $\mathrm{Var}(\eta_j) \approx 4226$ and a standard deviation of roughly 65.0.

Against that noise scale, the decision threshold $q/4 = 16384$ is enormous. Even a very loose subgaussian or bounded difference tail argument yields an astronomically small bit error probability. Applying a union bound over 256 bits still leaves a negligible correctness failure probability. This does not constitute a tight proof, but it does justify the empirical observation that the benchmark and self diagnostic runs complete with exact file recovery.

The implementation now also records decapsulation correctness diagnostics in benchmark schema v3. Across the uploaded formal benchmark corpus used in this manuscript, the JSON artifacts report 1670 out of 1670 valid decapsulation correctness checks and zero invalid ones. This is empirical evidence only. It does not replace a proof tuned to the exact structured secret law, but it does show that the present implementation behaves consistently in the tested regime.

The FO layer then transforms low bit error probability into KEM correctness in the usual way. If $\mu' = \mu$, the reconstructed ciphertext and tag match and decapsulation returns the real shared key. If any bit is decoded incorrectly, the recomputed transcript will differ and the constant time selection returns the fake key. From an external viewpoint this keeps failure handling pseudorandom and avoids obvious decapsulation side channel leakage through branch structure.

The revised manuscript does not claim novel decoding theory. Its contribution is simply to spell out that the fixed weight formal miner makes the variance accounting cleaner than before, because the dominant inner product noise term now has a known size rather than lying only inside a wide popcount window.

\begin{align*}
\eta_j &= \sum_{i \in I_j} e_i + e_2^{(j)} - \langle e_1^{(j)}, s \rangle, \\
\mathrm{Var}(\eta_j) &\approx 64 \cdot 2 + 2 + 2048 \cdot 2 = 4226, \qquad
\mathrm{SD}(\eta_j) \approx 65.0.
\end{align*}

\section{Security discussion and scope of claims}
The security claim of HyperFrog remains intentionally narrow. The scheme is an experimental unstructured LWE style KEM with a structured binary secret law. No claim is made that graph cycle rank itself is a cryptographic hard problem, and no claim is made that the structured secret law is as well studied as the distributions used in standardized KEMs. The right conservative reading is that if LWE remains hard for this secret family, and if the usual random oracle model assumptions hold for the FO transform, then the construction has a plausible experimental basis.

Two assumptions are particularly delicate. The first is the standard pseudorandom matrix expansion assumption for ChaCha20 derived rows. The second is the transfer of LWE hardness from more familiar small or binary secret families to the present connected graph, exact weight, high cycle rank conditioned family. The literature does contain substantial analysis of binary and sparse secrets, but the graph conditioned correlations introduced here remain an open problem.

The revision actually sharpens the risk statement rather than weakening it. Because the formal miner now defines a clearly non uniform law, the paper no longer hides behind the suggestion that the secret may be almost uniform under the published parameters. Instead it states plainly that the secret distribution is structured, that this structure is intentional, and that its security consequences require dedicated study. This is scientifically more honest and, in my view, more useful.

Potential attack surfaces include statistical distinguishing attacks on $As + e$ when $s$ is restricted to a graph conditioned family, correlation attacks that exploit frontier growth artifacts, and any algorithm that learns nontrivial linear information from the fixed occupied weight together with local adjacency constraints. Because the secret is not a product Bernoulli source, classical entropy calculations alone are not enough to justify security. One needs structural analysis of the family itself.

The FO transform side is more conventional. Provided the core encryption hides $\mu$ under the stated assumptions, the deterministic reencryption and transcript binding tag follow the usual route to chosen ciphertext security in the random oracle model. The code derives the reencryption seed from the public key digest and candidate $\mu'$, recomputes the ciphertext deterministically, compares $u$, $v$, and tag in constant time, and then selects between real and fake keys without secret dependent branching. This part is conceptually standard even though the sparse row subset encryption rule is not identical to FrodoKEM.

Side channel scope should also be stated conservatively. The implementation improves constant time behavior in one important place by using a branchless distance to $q/2$ test during decapsulation. It also uses secure memory helpers for shared keys and stricter parsing for serialized inputs. None of this is a substitute for a full side channel audit. In particular, large vector arithmetic, memory traffic during row expansion, and password based key unlock should still be treated as audit targets.

The revised paper therefore offers a bounded claim. HyperFrog is a transparent experimental system that now states its formal secret law honestly and reports its benchmark semantics clearly. Whether that secret law is cryptographically robust remains an open research question and a natural target for future public analysis.

\section{Implementation details in \texttt{hyperfrog36.cpp}}
The reference implementation accompanying this revision is \texttt{hyperfrog36.cpp}, aligned here with the v36.0 code line and benchmark schema v3. The public repository is available at \url{https://github.com/victormeloasm/HyperFrog} and should be treated as the canonical implementation artifact for this manuscript~\cite{hyperfrogrepo}. Several changes are directly relevant to the claims of this manuscript because they affect either the stated secret law or the interpretation of the benchmark artifacts.

The first implementation level change is the explicit separation of miner modes. The code path named \texttt{formal} calls \texttt{hf\_mine\_shape\_reference}, which implements the connected frontier growth process and returns failure on timeout. The code path named \texttt{practical} calls \texttt{hf\_mine\_shape\_practical}, which keeps the largest component of a random bitset and may return a deterministic connected fallback shape. Because these two paths define different output laws, the manuscript refers to them separately throughout.

The second change is the renaming of the topological statistic in the code itself. The structure \texttt{HFTopo} now exposes the field \texttt{cycle\_rank} instead of using the overloaded word \texttt{genus}. The topology routine counts vertices, face adjacency edges, and connected components, then stores \texttt{cycle\_rank = E - V + C}. This closes the gap between the language of the paper and the behavior of the code.

The third change is in decapsulation. The branchless helper \texttt{ct\_is\_close\_mask} computes the modular distance of a value to $q/2$ and returns a mask indicating whether the candidate lies in the middle half of $\mathbb{Z}_q$. Each message bit is reconstructed from that mask without a secret dependent branch. The benchmark harness also records a decapsulation correctness object that tracks whether the diagnostic decapsulation passed and records margins inside the decode bands.

The fourth change is the benchmark schema itself. Schema v3 records per sample key generation outcomes, timing, and mining diagnostics. In the formal miner this includes attempts, accepted count, accepted cycle rank, connected component count, and shape weight. One field deserves a caveat. The current JSON field named \texttt{acceptance\_probability} is the ratio \texttt{accepted / attempts} inside the parallel search loop. Because the loop stops once one acceptable sample is found, this field is useful for implementation introspection but it is not yet a calibrated acceptance probability of the formal law. This paper therefore treats it as an instrumentation field rather than as a theorem level quantity.

The fifth change is stricter serialization and file handling. The code checks magic bytes, version identifiers, lengths, and exact reads before deserializing public keys, secret keys, and ciphertext files. The secret key parser also validates the AEAD identifier more strictly than before. These changes matter because benchmark harnesses that preserve audit artifacts are only useful if the artifact format itself is parsed defensively.

The sixth change is the benchmark split. The code has helper paths that can encrypt with a preloaded public key and decrypt with a preloaded secret key. This lets the benchmark measure KEM encapsulation, KEM decapsulation, file encryption, secret key unlock, file decryption after unlock, and total file decryption separately. Without this split one can easily misread Argon2id cost as cryptographic decryption cost.

The implementation also uses AVX2 optimized dot products when available, OpenMP parallelism when available, OpenSSL SHA3, libsodium \texttt{randombytes}, and either AES 256 GCM or XChaCha20 Poly1305 for secret key wrapping depending on platform capability. These details do not change the mathematical design, but they do explain the benchmark profiles reported later in the paper.

\section{Experimental methodology}
The benchmark artifacts supplied with this revision come from a small corpus rather than from one single timing run. The corpus is intended to answer four questions. First, does the formal miner behave stably and produce the exact connected weight constrained secret family claimed in the manuscript. Second, what is the latency of the KEM itself under several practical thread settings. Third, how much of file decryption time comes from the cryptographic path and how much comes from password based secret key unlock. Fourth, how sharply do the presently chosen threshold and budget sweeps actually stress the miner.

\begin{table}[H]
\centering
\caption{Benchmark suites used in this revision.}
\small
\setlength{\tabcolsep}{4pt}
\renewcommand{\arraystretch}{1.12}
\begin{tabularx}{\textwidth}{>{\raggedright\arraybackslash}p{0.23\textwidth} >{\centering\arraybackslash}p{0.10\textwidth} >{\centering\arraybackslash}p{0.12\textwidth} >{\centering\arraybackslash}p{0.10\textwidth} >{\centering\arraybackslash}p{0.11\textwidth} >{\raggedright\arraybackslash}X}
\toprule
\textbf{Suite} & \textbf{Mode} & \textbf{KEM iters} & \textbf{Threads} & \textbf{File size} & \textbf{Purpose} \\
\midrule
\texttt{bench\_light} & formal & 20 & 8 & 4 MiB & quick sanity benchmark \\
\texttt{bench\_mid} & formal & 100 & 16 & 16 MiB & medium throughput run \\
\texttt{bench\_heavy} & formal & 500 & 32 & 128 MiB & large KEM corpus and large file run \\
\texttt{file256} & formal & 50 & 16 & 256 MiB & large file scaling run \\
\texttt{cmp\_formal} & formal & 200 & 16 & 64 MiB & formal comparison baseline \\
\texttt{cmp\_practical} & practical & 200 & 16 & 64 MiB & engineering comparison baseline \\
\shortstack[l]{\texttt{th1, th8,}\\\texttt{th16, th32}} & formal & 100 each & varied & 16 MiB & thread sweep \\
\shortstack[l]{\texttt{rank6, rank8,}\\\texttt{rank10, rank12}} & formal & 50 each & 16 & 8 MiB & threshold sweep \\
\shortstack[l]{\texttt{mine\_1000, \ldots,}\\\texttt{mine\_30000}} & formal & 50 each & 16 & 8 MiB & budget sweep \\
\bottomrule
\end{tabularx}
\normalsize
\end{table}

Across the formal suites listed above, the uploaded artifacts contain 1670 formal key generation samples and 1670 decapsulation correctness trials. All formal file integrity checks in the supplied JSON objects report exact recovery. The practical corpus adds 200 further engineering mode samples for comparison but is not used to support formal distribution claims.

KEM latency is measured separately for encapsulation and decapsulation. File processing is measured in four ways. The first number is file encryption time. The second is secret key unlock time, which is zero when the secret key is stored without password protection and positive when Argon2id based wrapping is used. The third is cryptographic file decryption time after a usable secret key is already in memory. The fourth is total end to end file decryption time, which is the sum of unlock and cryptographic decryption.

This methodology matters because password based secret key storage is an implementation level protection for long term secret files, not an inherent cost of the LWE KEM or the AEAD file path. Reporting only the total decrypt time would therefore mix a memory hard password derivation cost with the actual cost of decryption.

One further methodological point is important for honesty. The current threshold sweep uses $\tau \in \{6,8,10,12\}$ and the current budget sweep uses 1000 ms to 30000 ms. Because the observed accepted cycle ranks in the formal runs already cluster around roughly 2600 to 2800 and the observed key generation times are typically below 1 ms, these sweeps do not yet probe a difficult acceptance regime. They are still useful as stability checks, but they should not be misread as a calibrated acceptance study.

\section{Benchmark results}
The benchmark corpus supports three implementation claims strongly and one calibration claim only weakly. First, the formal miner is stable in the tested regime. Second, the practical miner is clearly distinct from the formal one in its observed geometry. Third, once the secret key has been unlocked, the file encryption and decryption paths are comparatively balanced. The weak point is calibration of threshold and budget stress, which remains for future work.

\begin{table}[H]
\centering
\caption{Formal benchmark corpus summary across all uploaded formal JSON artifacts.}
\renewcommand{\arraystretch}{1.18}
\begin{tabularx}{\textwidth}{>{\raggedright\arraybackslash}p{0.43\textwidth} >{\centering\arraybackslash}p{0.18\textwidth} >{\raggedright\arraybackslash}X}
\toprule
\textbf{Metric} & \textbf{Value} & \textbf{Comment} \\
\midrule
Formal key generation samples & 1670 & Aggregate across light, medium, heavy, file scaling, thread, threshold, and budget suites. \\
Formal key generation successes & 1670 & No observed formal key generation failure in the uploaded corpus. \\
Formal key generation timeouts & 0 & The present corpus does not stress a timeout regime. \\
Observed occupied weight & 2048 exactly & Every formal sample has exact occupied weight 2048. \\
Observed connected components & 1 exactly & Every formal sample is connected. \\
Observed cycle rank range & 2315 to 2856 & Aggregate over all formal samples in the uploaded corpus. \\
Observed cycle rank mean & 2687.3 & Aggregate mean over the formal corpus. \\
Decapsulation correctness diagnostics & 1670 valid, 0 invalid & Empirical evidence only, not a proof. \\
\bottomrule
\end{tabularx}
\end{table}

\begin{figure}[H]
\centering
\includegraphics[width=0.88\textwidth]{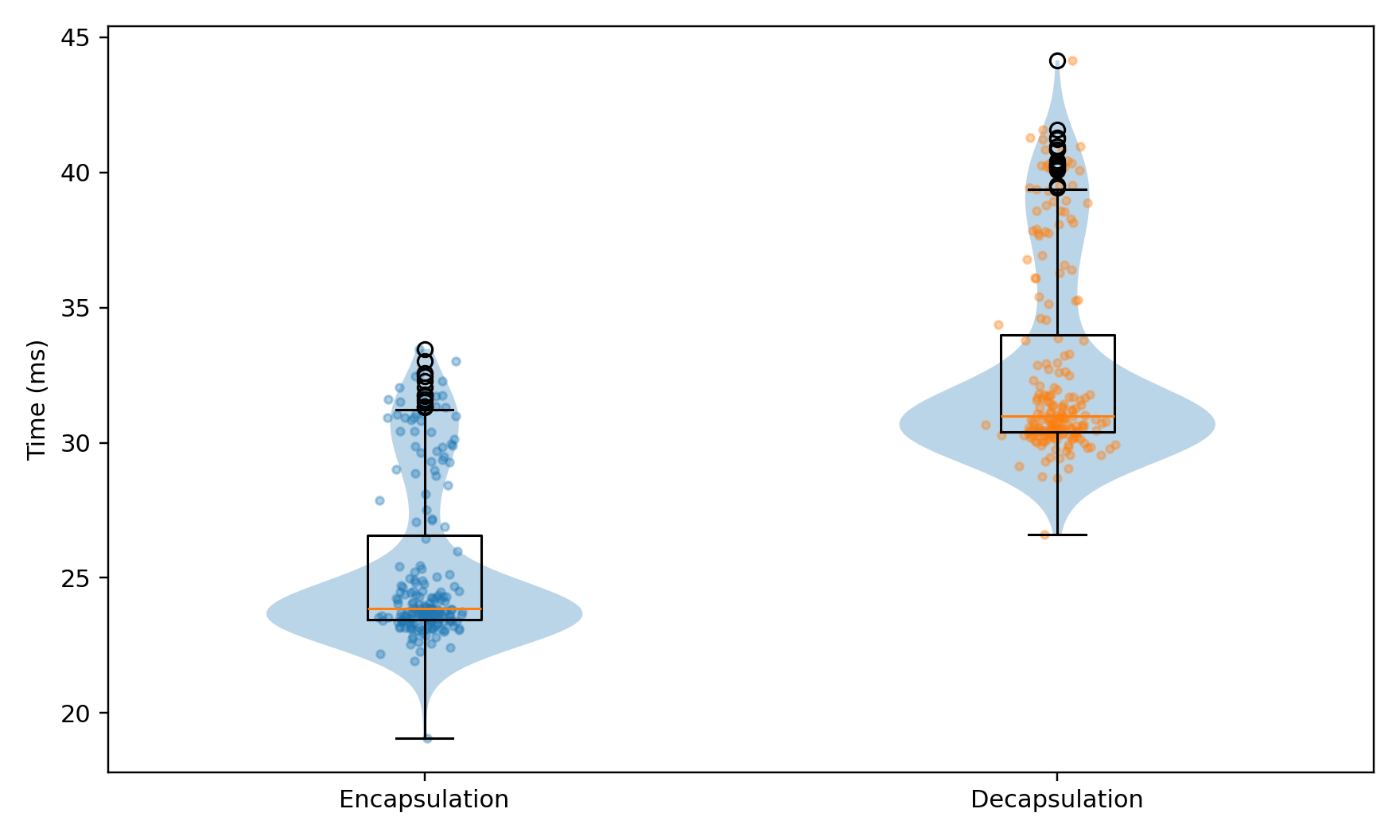}
\caption{Representative formal KEM latency distribution from the 16 thread, 200 iteration comparison run. Decapsulation is consistently slower than encapsulation, but the spread remains compact.}
\label{fig:kemlat}
\end{figure}

Figure~\ref{fig:kemlat} shows a representative KEM latency distribution from the 16 thread formal comparison run with 200 iterations. In that run the mean encapsulation time is 25.3406 ms and the mean decapsulation time is 32.7150 ms. The spread is modest, and decapsulation is consistently slower than encapsulation, which is expected because it reconstructs candidate message bits, recomputes the deterministic encapsulation transcript, and performs constant time validation.

The thread sweep shows that more threads do not monotonically improve the observed benchmark profile in the current regime. In the uploaded throughput suite, 16 threads gave the best observed KEM latencies, while 32 threads incurred visible overhead. This does not mean that OpenMP is useless. It means only that for this miner and this machine, the present workload is light enough that thread management overhead becomes significant at 32 threads.

\begin{table}[H]
\centering
\caption{Thread sweep summary from the formal benchmark corpus.}
\renewcommand{\arraystretch}{1.18}
\begin{tabular}{ccccc}
\toprule
\textbf{Threads} & \textbf{KeyGen mean ms} & \textbf{KeyGen P95 ms} & \textbf{Encap mean ms} & \textbf{Decap mean ms} \\
\midrule
1 & 0.19 & 1.00 & 152.9700 & 158.4720 \\
8 & 0.27 & 1.00 & 26.7992 & 31.8745 \\
16 & 0.19 & 1.00 & 19.1723 & 25.0365 \\
32 & 0.54 & 1.00 & 26.7488 & 35.5154 \\
\bottomrule
\end{tabular}
\end{table}

\begin{figure}[H]
\centering
\includegraphics[width=0.88\textwidth]{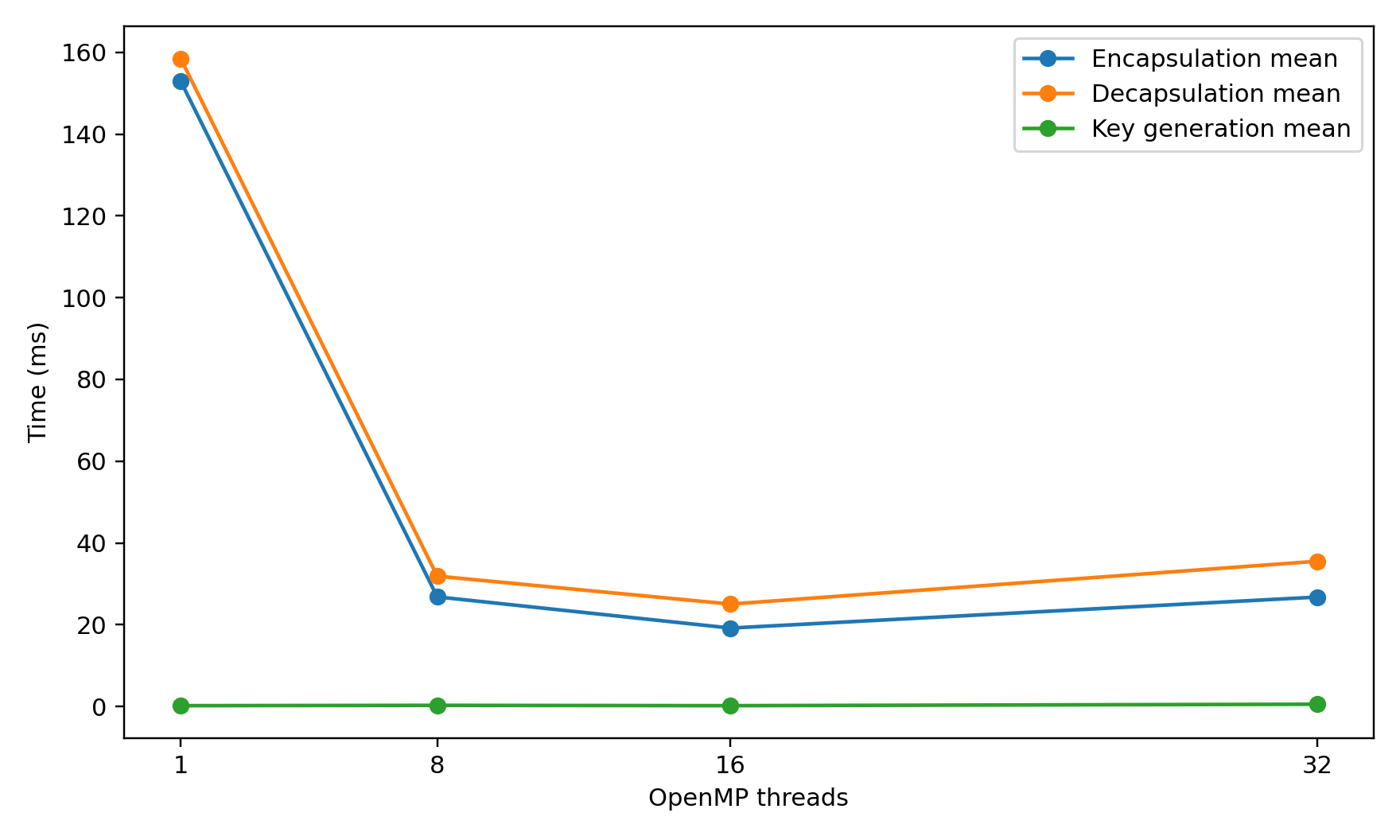}
\caption{Observed mean timings in the formal thread sweep. The present miner is so light in this regime that 32 threads add overhead rather than improving performance.}
\label{fig:threads}
\end{figure}

The formal versus practical comparison is also informative. The formal miner keeps exact weight 2048 in every sample and produces observed cycle ranks centered near 2692 in the comparison run. The practical miner, by contrast, produces weights centered near 2030 and cycle ranks centered near 1000. This is a large geometric difference and it is one reason the manuscript keeps the two modes separate.

\begin{table}[H]
\centering
\caption{Formal versus practical comparison from the 200 iteration, 16 thread benchmark pair.}
\renewcommand{\arraystretch}{1.18}
\begin{tabularx}{\textwidth}{>{\raggedright\arraybackslash}p{0.18\textwidth} >{\centering\arraybackslash}p{0.12\textwidth} >{\centering\arraybackslash}p{0.15\textwidth} >{\centering\arraybackslash}p{0.22\textwidth} >{\centering\arraybackslash}p{0.13\textwidth} >{\centering\arraybackslash}p{0.13\textwidth}}
\toprule
\textbf{Mode} & \textbf{KeyGen mean ms} & \textbf{Weight mean} & \textbf{Weight range} & \textbf{Cycle rank mean} & \textbf{Cycle rank range} \\
\midrule
Formal & 0.435 & 2048.0 & 2048 to 2048 & 2692.13 & 2556 to 2833 \\
Practical & 0.580 & 2030.0 & 1960 to 2133 & 1000.02 & 913 to 1108 \\
\bottomrule
\end{tabularx}
\end{table}

\begin{figure}[H]
\centering
\includegraphics[width=0.82\textwidth]{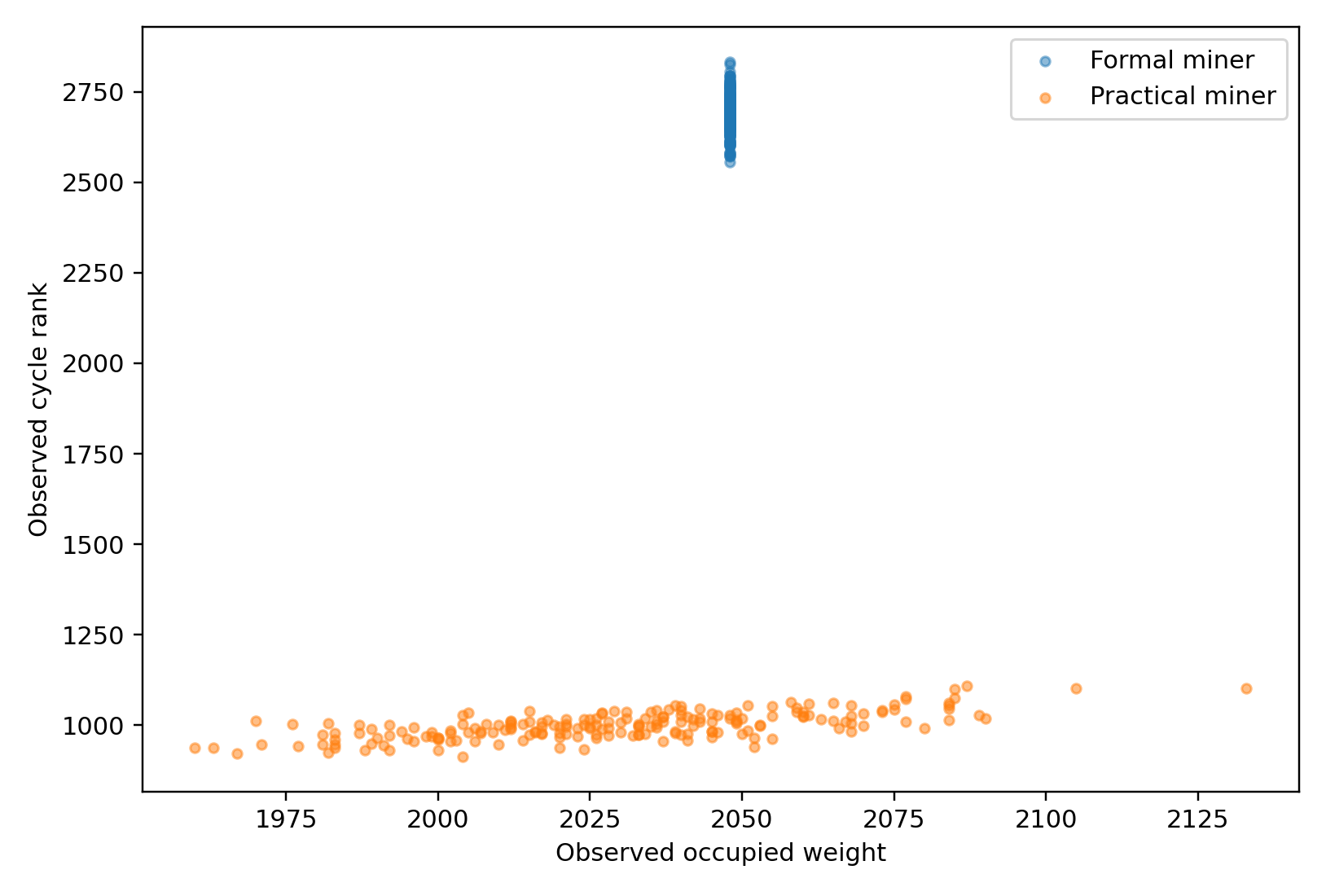}
\caption{Observed geometry in the formal and practical comparison runs. The formal miner remains on exact weight 2048 with much higher cycle rank, while the practical miner explores a distinctly different region.}
\label{fig:modecompare}
\end{figure}

\begin{figure}[H]
\centering
\includegraphics[width=0.92\textwidth]{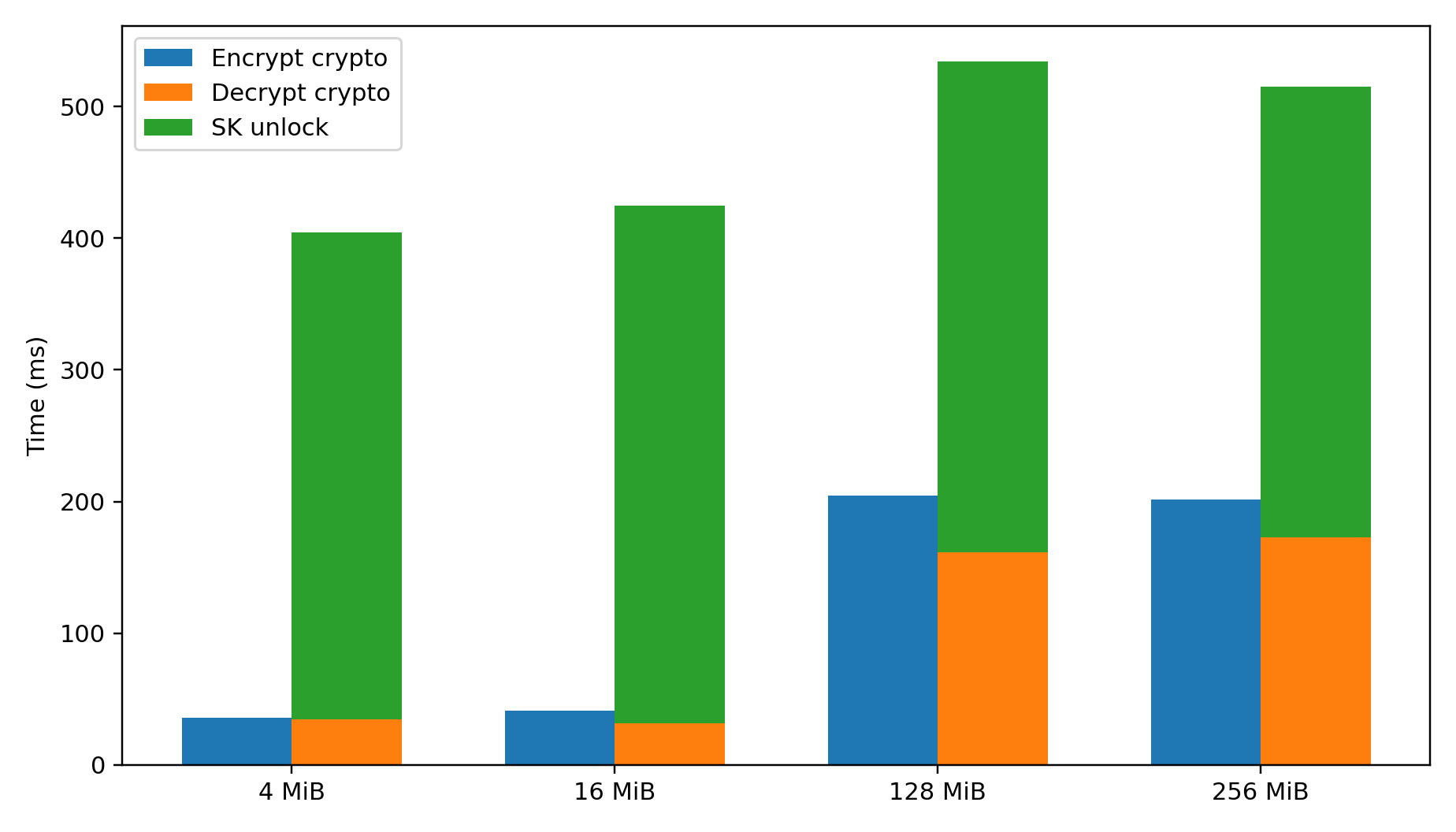}
\caption{File processing breakdown across the uploaded formal file benchmarks. The dominant fixed cost is password based secret key unlock. The cryptographic file path then scales with file size.}
\label{fig:filescale}
\end{figure}

File processing numbers tell a different story from KEM latency. Once the secret key is already available in memory, the cryptographic file path is comparatively balanced. The main additional delay comes from password based secret key unlock. Figure~\ref{fig:filescale} shows this clearly across 4 MiB, 16 MiB, 128 MiB, and 256 MiB file runs. The unlock cost is large and comparatively flat. The cryptographic file path scales with file size as expected, though the measured throughputs are not perfectly monotone, which is unsurprising in a small benchmark corpus that includes filesystem and caching effects.

The threshold and budget sweeps are honest but weak. In the current uploaded regime, increasing the target cycle rank from 6 to 12 changes the observed key generation mean only slightly, and varying the mining budget from 1000 ms to 30000 ms does not produce timeouts. This is evidence that the present thresholds and budgets are non binding. It is not evidence that the formal miner has already been calibrated at a scientifically meaningful acceptance boundary.

\section{Discussion of the benchmark numbers}
The quantitative picture that emerges from the revised corpus is easier to interpret than in earlier drafts because the benchmark semantics are now explicit and the corpus is broader.

The strongest empirical claim is stability of the formal miner in the uploaded regime. Across 1670 formal samples, the miner always produced exact weight 2048 and one connected component, and the observed accepted cycle ranks stayed in a relatively tight band compared with the full graph state space. That is good evidence that the connected growth law is stable as implemented.

The second strong claim is separation of formal and practical modes. The practical miner remains useful for engineering tasks, but its observed geometry is plainly different. Because it produces lower cycle ranks and variable occupied weight, it cannot stand in for the formal law in a security discussion. The benchmark corpus makes this distinction visible rather than merely asserting it.

The third strong claim is the interpretation of file latency. For the supplied password protected secret keys, most user visible decrypt time comes from local Argon2id based secret key unlock rather than from the cryptographic decrypt path. That is an important engineering conclusion because one should not compare HyperFrog total decrypt time against another prototype's crypto only decrypt time without careful normalization.

The weaker claim concerns miner calibration. The current threshold sweep and budget sweep do not yet create a hard acceptance regime. In the present corpus, the accepted cycle rank is already far above the requested threshold, and the observed key generation time is already far below the requested budget. The current sweeps therefore support a statement about robustness, not yet a statement about calibrated acceptance probability.

\section{Limitations and open problems}
The first limitation is conceptual. No reduction currently shows that unstructured LWE with this connected graph, exact weight secret family is as hard as standard binary secret LWE. This remains the main cryptanalytic question raised by the design.

The second limitation is bandwidth. The ciphertext is still about 2.10 MB because the present design stores one full $u$ vector per message bit. Any attempt to move HyperFrog closer to practical deployment would need a much more compact ciphertext representation, perhaps by batching, compressing, or redesigning the message embedding step. The present paper does not solve that problem.

The third limitation is miner calibration. The current artifact suite does not yet probe a difficult threshold regime. The uploaded threshold sweep uses small target values and the uploaded budget sweep uses generous budgets relative to observed key generation time. A future study should move the threshold closer to the empirically observed cycle rank band and move the budget toward the millisecond scale where timeouts actually become possible.

The fourth limitation is acceptance semantics. The current JSON field named \texttt{acceptance\_probability} is useful for implementation introspection, but because it is tied to a stop on first success parallel search loop, it should not yet be interpreted as the acceptance probability of the formal law in a mathematical sense. That quantity needs a cleaner estimator and a dedicated study.

The fifth limitation is side channel scope. The implementation is more careful than before, but it has not been subjected to a full external side channel audit. This is especially relevant because the code mixes large vector operations, hash based derivation, file parsing, and password based secret key handling.

The sixth limitation is semantic maturity. HyperFrog is still best understood as a research platform for studying structured binary secret laws inside an unstructured LWE KEM, not as a candidate for deployment. The revision improves precision and honesty. It does not convert the scheme into a standardized production system.

\section{Conclusion}
This revision rewrites HyperFrog around four clarifications. The formal topological invariant is graph cycle rank $\beta_1$ of the occupied six neighbor graph. The formal miner is a connected frontier growth process with exact occupied weight, not a weak rejection sampler over nearly uniform occupancy fields. The benchmark methodology separates secret key unlock from the cryptographic decrypt path and records per sample mining and decapsulation diagnostics. The practical miner remains present in code, but it is explicitly separated from formal claims.

These changes do not make HyperFrog finished. What they do is make it much more interpretable. A reviewer can now read the manuscript, inspect the code, and know exactly which secret law is being claimed, which implementation mode is merely practical, which benchmark numbers belong to cryptography rather than password based local key protection, and which benchmark sweeps remain non binding.

The remaining challenge is now sharper and more interesting. Given this explicit connected graph, exact weight binary secret law, how far do known LWE analyses and attacks extend, and what structure specific attacks might exist. Answering that question would move HyperFrog from a corrected experimental construction toward a better understood research object.

\section*{Appendix A. Benchmark semantics and artifact interpretation}
The benchmark JSON used in this revision reports \texttt{encaps\_ms} and \texttt{decaps\_ms} for the KEM core only. These values do not include file processing. The file object then reports \texttt{enc\_ms}, \texttt{dec\_unlock\_ms}, \texttt{dec\_crypto\_ms}, and \texttt{dec\_total\_ms}. The relation is \texttt{dec\_total\_ms = dec\_unlock\_ms + dec\_crypto\_ms} up to measurement rounding.

The \texttt{integrity\_ok} field reports whether the decrypted file matches the original file exactly. In the uploaded corpus used here this value is true in every supplied JSON object. This matters because performance numbers without an accompanying integrity result are of limited scientific value in a research prototype.

The key generation object records sample counts, success counts, timeouts, distributions of elapsed time, and per sample mining notes. The current field \texttt{acceptance\_probability} should be interpreted carefully. In the present code line it is the ratio \texttt{accepted / attempts} inside the search loop that stops once a valid sample is found. It is therefore an implementation diagnostic and not yet a calibrated acceptance probability for the formal law.

The benchmark also records the parameter set, target cycle rank, thread count, mining budget, number of iterations, file size, and mode. These metadata items are part of the scientific result because they determine both the secret generation regime and the timing environment.

\section*{Appendix B. Reference implementation notes for reviewers}
Reviewers comparing the manuscript with \texttt{hyperfrog36.cpp} should focus on five code paths. The first is \texttt{hf\_mine\_shape\_reference}, which implements the formal miner. The second is \texttt{hf\_compute\_topology}, which computes vertices, edges, components, and cycle rank. The third is \texttt{hf\_decaps} together with \texttt{ct\_is\_close\_mask}, which implements branchless bit decoding and constant time real or fake key selection. The fourth is the benchmark harness, which separates secret key unlock from cryptographic file decryption and records per sample mining diagnostics. The fifth is the practical miner path, which should be read as engineering only and not as part of the formal specification.

Reviewers should also note one compatibility detail. Some command line aliases and older repository terminology may still exist for backward compatibility in the code line, but the manuscript consistently uses the terms \textit{cycle rank}, \textit{formal miner}, and \textit{practical miner}. Where the two differ, the manuscript terminology should be treated as authoritative.


\begin{thebibliography}{9}
\bibitem{regev2005}
O. Regev. On lattices, learning with errors, random linear codes, and cryptography. \textit{Proceedings of the 37th Annual ACM Symposium on Theory of Computing}, 2005.

\bibitem{peikert2016}
C. Peikert. A decade of lattice cryptography. \textit{Foundations and Trends in Theoretical Computer Science}, 2016.

\bibitem{micciancio2009}
D. Micciancio and O. Regev. Lattice based cryptography. In \textit{Post Quantum Cryptography}, Springer, 2009.

\bibitem{frodokem2020}
J. W. Bos et al. FrodoKEM: Learning with errors key encapsulation. NIST Post Quantum Cryptography Project submission, 2020.

\bibitem{fujisakiokamoto1999}
E. Fujisaki and T. Okamoto. Secure integration of asymmetric and symmetric encryption schemes. \textit{Advances in Cryptology, CRYPTO 1999}.

\bibitem{fips202}
National Institute of Standards and Technology. FIPS 202, SHA 3 Standard: Permutation Based Hash and Extendable Output Functions, 2015.

\bibitem{hyperfrogrepo}
Victor Duarte Melo. HyperFrog reference implementation and benchmark artifacts. GitHub repository, HyperFrog v36.0 branch and benchmark schema v3, March 2026. \url{https://github.com/victormeloasm/HyperFrog}.

\bibitem{melo2025vffnd}
Victor Duarte Melo. Scale Normalized Overlap and the Metric Geometry of Measurable Shapes in R n. Academia.edu research paper, 2025. \url{https://www.academia.edu/164977278/Scale_Normalized_Overlap_and_the_Metric_Geometry_of_Measurable_Shapes_in_R_n}.

\end{thebibliography}
\end{document}